# A Framework for Policy Crowdsourcing

JOHN PRPIĆ           - Beedie School of Business - Simon Fraser University
ARAZ TAEIHAGH[1]     - City Futures Research Centre - University of New South Wales
JAMES MELTON         - College of Business Administration - Central Michigan University

--------------------------------------------------------------------------------------------------------

**Abstract**

What is the state of the literature in respect to Crowdsourcing for policy making? This work attempts to answer this question by collecting, categorizing, and situating the extant research investigating Crowdsourcing for policy, within the broader Crowdsourcing literature. To do so, the work first extends the Crowdsourcing literature by introducing, defining, explaining, and using seven universal characteristics of all general Crowdsourcing techniques, to vividly draw-out the relative trade-offs of each mode of Crowdsourcing. From this beginning, the work systematically and explicitly weds the three types of Crowdsourcing to the stages of the Policy cycle as a method of situating the extant literature spanning both domains. Thereafter, we discuss the trends, highlighting the research gaps, and outline the overlaps in the research on Crowdsourcing for policy, stemming from our analysis.

**Keywords:** Crowdsourcing, Policy Cycle, Policy Processes, Policy Stages, Virtual Labor Markets, Tournament Crowdsourcing, Open Collaboration Crowdsourcing,

## 1. Introduction

Crowdsourcing (Howe 2006; 2008) involves organizations using IT to engage Crowds comprised of groups and individuals for the purpose of completing tasks, solving problems or generating ideas. In the last decade, many organizations have turned to Crowdsourcing to engage with consumers, accelerate their innovation cycles, and to find new ideas for their brands (Afuah & Tucci 2012; Bayus 2013; Brabham 2013). As Crowdsourcing has become an increasingly popular method for organizations to gather IT-mediated input from individuals, the phenomenon has also spread to non-commercial contexts too. Recently, Crowdsourcing (Brabham 2008) and the "wisdom of crowds" (Surowiecki 2005) are beginning to be applied to different aspects of policy making, as illustrated by nascent research in the transportation (Nash 2009) and urban planning domains (Seltzer & Mahmoudi 2013). Yet, despite the advancing use of Crowdsourcing in general, and its nascent application in policy contexts, to our knowledge research has yet to emerge that systematically investigates both domains simultaneously.

In this work, we begin to address this salient gap in our knowledge by first categorizing the extant research on Crowdsourcing for policy, within the framework of the policy cycle (Stone 1988; Howlett et al. 1995). In doing so, we serve to situate and organize the extant literature, merging both domains into a useful framework which illustrates the current trends, highlights the research gaps, and indicates the empirical approaches thus far employed to investigate policy Crowdsourcing.

---

[1] Corresponding author: araz.taeihagh@new.oxon.org, City Futures Research Centre, FBE, UNSW, Sydney 2052, NSW, Australia.

In the ensuing sections of this work, we introduce the three types of Crowdsourcing in Section #2, briefly detailing their similarities and differences along seven independent factors relevant to Crowdsourcing techniques. Thereafter, we introduce the policy cycle in Section #3 and discuss its use and application as a guiding framework for policy processes, and merge these two frameworks into a convenient table that organizes and situates the extant literature on policy Crowdsourcing, highlighting the apparent gaps and overlaps in the extant research. In Section #4 we discuss the ramifications of our study for the research and policy practitioner communities, before concluding with a summary of our contributions.

## 2. Crowdsourcing

Crowdsourcing is an IT-mediated problem solving, idea-generation, and production model that leverages the dispersed knowledge of groups and individuals to produce heterogeneous resources for organizations (Hayek, 1945; Brabham, 2008; Prpić & Shukla, 2013; 2014; Prpić, Shukla, Kietzmann & McCarthy, forthcoming). Problem-solving, idea-generation and production are "sourced from Crowds" through means such as micro-tasking (asking individuals to execute short tasks for pay), open collaboration (asking individuals to volunteer input online) or through tournament-based competitions (where individuals submit contributions in the hope of 'winning a prize'). As an overall approach to engaging dispersed knowledge through IT, Crowdsourcing processes serve to blend the efficiency and control of traditional, top-down managed processes, with the benefits of bottom-up open innovation and creativity (Brabham 2008; Howe 2006; Howe 2008).

Organizations can launch Crowdsourcing initiatives on their own in-house platforms, therein seeking to coalesce a proprietary Crowd, as Dell (IdeaStorm) and Starbucks (MyStarbucks) have, or an organization can commission Crowdsourcing intermediaries to provide the requisite IT means and a "built-in" Crowd, as a service for fees (Bayus 2013). Though still emerging and evolving in form and function, Crowdsourcing phenomena and the Crowdsourcing literature have recently begun to coalesce, where many scholars now view Crowdsourcing as occurring as three distinct forms of IT-mediated collaboration (Estellés-Arolas & González-Ladrón-de-Guevara 2012, de Vreede et al. 2013, Prpić, Jackson & Nguyen 2014). In the ensuing subsections we'll focus on each distinct type in turn.

### 2.1 Virtual Labor Marketplaces (VLM's)

A virtual labor marketplace (VLM) is an IT-mediated market for spot labor, typified by endeavors like Amazon's M-Turk and Crowdflower, where individuals and organizations can agree to execute work in exchange for monetary compensation. These endeavours are generally thought to exemplify the "production model" aspect of Crowdsourcing (Brabham 2008), where workers undertake microtasks for pay.

Microtasks, such as the translation of documents, labelling/tagging photos, and transcribing audio (Narula et al. 2011), are generally considered to represent forms of human computation (Michelucci 2013, Iperiotis & Paritosh 2011), where human intelligence is asked to undertake tasks that are not currently achievable through artificial intelligence. Though human computation tasks cannot be tackled by the most advanced forms of AI, they are rather mundane for all human intelligences to undertake.

The size of the overall Crowd available at any one time to undertake tasks at these VLM's is immense, where Crowdflower for example has over 5 million laborers available at any given time (see http://crowdflower.com). Therefore, microtasking through VLM's can be rapidly completed on a massively parallel scale. Further, the participants in these VLM Crowds always undertake tasks independent of one another. Hence, though task completion can be actuated on a massively parallel scale for organizations, these tasks are undertaken without collaboration of any sort amongst individual Crowd members. Similarly, teams do not form at VLM's to undertake tasks in groups. Altogether the literature (Prpić & Shukla 2013; 2014, Prpić, Shukla, Kietzmann & McCarthy 2015)has identified the episodic nature of VLM labourer contributions s as a key distinction amongst the different forms of Crowdsourcing.

**2.2 Tournament-Based Collaboration**
A separate form of Crowdsourcing is known as tournament-based collaborations (TBC). In tournament-based collaboration (TBC) organizations post their problems or opportunities to IT-mediated Crowds at web properties such as Innocentive, Eyeka, and Kaggle (Afuah and Tucci 2012) or through in-house platforms such as Challenge.gov (Brabham 2013b).

These web properties generally attract and maintain more or less specialized Crowds, premised upon the focus of the web property, though the properties are always open to new Crowd members joining. For example, the Crowd at Eyeka is coalesced around the creation of advertising collateral for brands, while the Crowd at Kaggle has formed around data science (Ben Taieb and Hyndman 2013, Roth and Kimani 2013). When applied to innovation, these platforms have been termed as innomediaries or open innovation platforms (Sawhney et al. 2003), and represent both the idea generation and problem solving aspects of Crowdsourcing (Morgan and Wang 2010, Brabham 2008).

The Crowd of participants at these sites is generally smaller when compared to the VLM's (for example, Kaggle has approximately a Crowd of 140,000 available -- see http://www.kaggle.com/solutions/connect), and the individual participants can choose not to be anonymous at these sites, in relation to their offline identities. Fixed amounts of prize money, and fixed amounts of prizes in total, are generally offered to the Crowd for the best solutions submitted, and can range from a few hundred dollars to a million dollars or more[2]. Some TBC's necessitate that their Crowds submit independent solutions to competitions (E.G. eYeka), while others such as TopCoder allow or even encourage team-formation and within-Crowd collaboration in competitions.

**2.3 Open Collaboration**
In an open collaboration (OC) model of Crowdsourcing, organizations post their problems/opportunities to the public at large through IT. Contributions from the Crowds in these endeavors are voluntary and do not generally entail monetary exchange. Posting on Reddit, starting an enterprise wiki (Jackson and Klobas 2013), or using social media (Kietzmann et al 2011) like Facebook and Twitter (Sutton et al 2014) to garner contributions, are prime examples of this type of Crowdsourcing. In this vein, Wikipedia is perhaps the most famous example of an OC application, "where a huge amount of individual contributions build solid and structured sources of data" (Prieur et al. 2008).

---
[2] (http://www.innocentive.com/files/node/casestudy/case-study-prize4life.pdf).

The scale of the Crowds available to these types of endeavors can vary significantly depending on the reach and engagement of the IT used, and the efficacy of the "open call" for volunteers. For example, as of May 2014, Reddit had approximately 2.8 million registered 'Redditors' (http://www.reddit.com/about) and though this mostly anonymous Crowd is quite large, there is little to guarantee the attention of any significant subset of the contributors when using Reddit.

It's important to note in respect to OC Crowdsourcing, that the Crowds in these endeavours are much less constrained in respect to self-organization (Prpić & Shukla 2013), relative to the other two types of Crowdsourcing. What this means for organizations is that the individuals in these OC Crowds, by virtue of their ready access to the same tools that the organizations are using, have, through their own volition, the ability to alter or amplify the agenda of organizational OC Crowdsourcing, through their own personal IT-mediated networks.

**2.4 Comparison of Crowdsourcing Techniques**

Prpić, Taeihagh & Melton (2014) compared the three types of Crowdsourcing discussed above across three universal dimensions--cost to implement, anonymity of individuals in the Crowd, and scale of the Crowd--using three-point estimates for each characteristic where possible. Here we build upon that work by extending the comparison to include four additional common characteristics to illustrate the stable and relative differences between the different forms of Crowdsourcing more vividly and comprehensively. This set of characteristics reflects a minimum and general consensus extracted from the literature (Prpić & Shukla 2013; 2014, Prpić, Shukla, Kietzmann & McCarthy 2015, de Vreede et al. 2013, Estellés-Arolas & Ladrón-de-Guevara 2012) and does not represent either an exhaustive set of characteristics, nor perfectly independent categorization, since many Crowdsourcing applications are hybrids of a sort, known to mix elements and features of the "pure play" forms that are thus far known, and that have thus far emerged.

The dimensions used for comparison of all Crowdsourcing techniques are listed below and discussed in turn.

    1) Cost
    2) Anonymity
    3) Scale of Crowd
    4) IT Structure
    5) Time required to implement
    6) Task Magnitude
    7) Reliability of the Crowd

**2.4.1 Cost**

The cost dimension refers to the typical cash outflows for an organization when implementing a Crowdsourcing technique for any purpose. Open Collaboration Crowdsourcing techniques, such as the use of Twitter, Reddit, Facebook, or a Wiki, are essentially free for the implementer in terms of direct cash outlays, while TBC's and VLM's necessitate explicit cash outlays for their use.

TBC's are generally fixed-cost cash outlays, where the organization sets the number of prizes in a tournament, and the value per prize, ahead of launching the competition. VLM's on the other hand, necessitate variable-cost cash outlays on a per completed task basis. Since TBC's (with very few exceptions such as Dell's IdeaStorm and Mystarbucks idea, see Bayus 2012) and VLM's are generally accessed through 3rd party intermediation services, cash outflows are also necessitated to compensate the service provider. In addition, there are indirect and opportunity costs involved with every organizational action.

On a related note, there may also be a different calculus of "residual value" to each form of Crowdsourcing endeavour. So for example, it may be that building a Twitter Crowd for Open Collaboration may provide multiple and continuing benefits for an organization beyond any particular one-off project, whereas there may be less (or no) residual value to using VLM's or TBC's, given that the Crowds accessed are proprietary to the intermediary firm. Similarly, the internal organizational use of the three types of Crowdsourcing, with one's own employees for example, should also alter the fundamental cost/benefit calculus discussed here. We outline these interesting and important differences as potentially fruitful avenues for future Crowdsourcing research.

### 2.4.2 Anonymity

The anonymity dimension of our comparison refers to whether the participants in the Crowds at each of the three generalized Crowdsourcing types are anonymous (or not) in respect to their offline identity. In some cases, predominantly in forms of Open Collaboration (such as in Google+, Facebook, Twitter or enterprise Wikis), the online and offline identity are identical. Whereas in VLM's there is essentially a form of 'methodological anonymity' found in all the intermediary platforms providing these services, where Crowd-workers are identified only by unique numeric identifiers (see Lease et al 2013, for an important exception). At the TBC intermediaries, anonymity is 'medium" in our estimation, since generally these platforms do not necessitate a concordance of online and offline identity, though some give Crowd-members the choice to use a pseudonym or their offline identity. At some TBC's such as Kaggle or Innocentive, there may actually be strong incentives for high-performers to eschew anonymity, so that their excellent performances can bolster their offline career prospects.

Taken altogether, the relative anonymity of Crowd-participants is important since anonymity is one method of maintaining privacy. Those organizations that are concerned with maintaining privacy for legal, ethical or moral reasons (I.E. Researchers & Health care organizations) or mandated to do so (I.E. Government institutions) will need to consider the liability that the different forms of Crowdsourcing anonymity supply.

### 2.4.3 Scale of Crowd Size

The size of Crowds available to organizations implementing one of the generalized forms of Crowdsourcing varies for each form. VLM's like Crowdflower boast 5 million plus members available at any one time for an organization to access, while the most successful TBC's like Kaggle, eYeka, and Innocentive boast Crowds of hundreds of thousands of members each. In effect, there may be something approaching or surpassing an order of magnitude of difference between VLM and TBC Crowds in general.

In respect to Open Collaboration, the situation varies significantly where applications like Twitter and Facebook have 100's of millions of members forming an upper limit of Crowd-size at these platforms (an order or magnitude or more than the largest VLM's), and other OC's such as Reddit count on a potential maximum Crowd in the single-digit millions. Further, other forms of open collaboration such as enterprise wikis, will of course vary only to the extent of the size of the firm itself, being implemented similarly with handfuls of individuals forming a Crowd in SME's, to hundreds of thousands of individuals in the largest firms.

The scale of Crowd differences outlined here are important for organizations to understand when choosing the form of Crowdsourcing to implement, since the scale of the Crowd represents a maximum limit to the number of potential contributions gathered from a Crowd, and thus as a potential constraint to the speed by which the needed contributions can be gathered.

### 2.4.4 IT Structure

The IT structure of the Crowdsourcing type can be found in either Episodic or Collaborative form premised upon the interface of the IT used to engage a Crowd. Episodic forms of Crowd-IT do not necessitate that Crowd members interact with one another through the IT, for resources to be derived from the Crowd, whereas, in Collaborative forms of Crowd-IT, the reverse is true, and Crowd participants must participate with one another through the IT for the organization to derive resources from the Crowd. This distinction is crucial given that all forms of Crowdsourcing are IT-mediated phenomena. Said another way, in Episodic forms of IT, social capital does not need to exist, be created, or maintained through the IT for Crowd-derived resources to be created. And the reverse is true for Collaborative forms of IT.

In our comparison of Crowdsourcing types, we find that VLM's are found to use Episodic IT structures, OC's are found to generally use Collaborative IT structures, while TBC's vary in this respect, where examples or elements of both forms of IT structure can be found to exist. Organizations considering to implement a type of Crowdsourcing, or to build their own, must very seriously consider these matters, given that the IT structure determines both the interaction between the organization and the Crowd, and the potential interaction of the Crowd-participants with each other.

### 2.4.5 Time to Implement the form of Crowdsourcing

The time required to implement a particular Crowdsourcing technique varies considerably amongst the available options. In our estimation VLM's necessitate the least amount of lead time to begin gathering Crowd-contributions, given the vast amount of on-demand labour available at all times at these platforms, and that one can join a VLM, and begin receiving contributions from the Crowd within minutes thereafter.

On the other hand, using a TBC like eYeka, Innocentive or Kaggle is a much more involved process, where the sponsoring firm generally works with the competition hosting intermediary to design the contest, and the prize and money distribution before the any Crowd members become involved. Further, competitions necessitate choosing an appropriate duration for the contest itself, generally ranging from a couple of week's time to many months or more. Similarly, at the end of the contest the winning submissions must be chosen from a sometimes vast range of competition entries received.

Open Collaboration on the other hand, varies in this respect, premised upon whether an organization already has the collaborative system in place or not. For example, if a firm has invested in developing a highly followed Twitter or Facebook network over time, then the time to receive Crowd contributions can be almost immediate. On the other hand, if such a presence needs to be built from scratch, the time to contribution from the Crowd can be very long indeed. Further, given that OC contributions are voluntary in nature, there is no guarantee that contributions ever manifest at all.

### 2.4.6 Task Magnitude

Task magnitude refers to the size and complexity of the tasks asked of, and received from the Crowd. VLM's are generally considered to elicit forms of human computation found in the magnitude of microtasks, TBC's are generally thought to elicit complete solutions to specific problems posed, while OC's can traverse the entire spectrum from microtasks to complete solutions, depending on a particular implementation. Task magnitude is an important consideration for organizations both within a form of Crowdsourcing, and amongst the three forms taken together. Essentially, task size is a function of the problem being posed to a Crowd.

In VLM's task size considerations are important given the massive parallel scale with which tasks can be undertaken, and because it is necessary to place an attractive price for each task into the market to attract Crowd-members to execute them. On the other hand, in TBC's task magnitude is essentially offered at the solution-level. In other words, organizations ask for, and receive fully-formed solutions to the entire problems that they offer up for the competition. Organizations must still choose the winning entries from the entries that are received, though in this case they are choosing amongst full-solutions, rather than aggregating individual task components into solutions, as with a VLM. In respect to OC task magnitude, broad variation exists spanning the spectrum from microtask to complete solutions. If an organization for example uses Twitter or Facebook to gather ideas from a Crowd, such Crowd-inputs would be closer to a microtask in a VLM, especially given the limitations of these platforms (ie 140 characters in Twitter), when detailed elaboration is desired. On the other hand, the use of an enterprise wiki in an organization is expected to accrue and evolve over time through Crowd contributions, and may approach something resembling a permanent, yet adjustable knowledge repository with relatively complete solutions. In all cases of Crowdsourcing, an organization will need to undertake some pre-task preparation, and some form of post-task processing of Crowd contributions, in order to generate the desired value, and task magnitude is central to these concerns.

### 2.4.7 Reliability of the Crowd

The reliability of the Crowd refers to the general consistency of each form of Crowdsourcing to supply the desired inputs for an organization. TBC's are considered to provide the highest level of reliability, given that said solutions are something approaching complete, once chosen. VLM's are considered to be of "medium" reliability given that these Crowds are available on-demand just like TBC's, though somewhat less reliable given the uncertainty around the price/time equation in the market.

In OC's, once more we see variability along this dimension relative to the specifics of the form of OC chosen. For example, if an organization uses Reddit, though there are approximately 3 million Redditors at the time of this writing, there is little to guarantee that any significant subset of Redditors will engage with the organization's effort. On the other hand in respect to an enterprise wiki for example, an organization, perhaps through directive, may be able to readily enforce or incentivize the entire internal employee population base to participate in short order, therefore considerably increasing the reliability of the Crowd.

**2.4.8 Summary**

Our aim with Section 2.2 was to provide a relative comparison of the three modes of Crowdsourcing in generalized form by unpacking the modes amongst the universal characteristics identified (see Table #1 below for the summary of the characteristics). In doing so, we highlighted some generalized trade-offs that face organizations in their implementation of Crowdsourcing techniques for any purpose. Our categorizations and assigned values are surely not exhaustive nor definitive, rather we feel that they are solid starting point, and useful extension to the extant Crowdsourcing literature in this respect.

In the ensuing section #3, we extend these classifications specifically to the Policy Cycle context to merge the modes of Crowdsourcing with the distinct stages of the Policy Cycle.

Table #1 – Comparison of Common Characteristics of Crowdsourcing Techniques

| Common Characteristics --------- Modes of Crowdsourcing | Cost of Implementing the Crowdsourcing technique | Anonymity of Individuals in the Crowd | Scale of Crowd Size | IT Structure of Crowdsourcing technique | Time required to implement Crowdsourcing technique | Complexity of task inputs from the Crowd | Reliability of the Crowd to provide desired inputs |
|---|---|---|---|---|---|---|---|
| Virtual Labor-Markets | Variable | High | High | Episodic | Low | Simple | Medium |
| Tournament-Based Collaboration | Fixed | Medium | Medium | Variable | Medium | Complex | High |
| Open Collaboration | Free | Variable | Variable | Collaborative | Variable | Variable | Variable |

### 3. The Policy Cycle

Jenkins (1978) defined public policy as "a set of interrelated decisions taken by a political actor or group of actors concerning the selection of goals and the means of achieving them within a specified situation where those decisions should, in principle, be within the power of those actors to achieve". As such a policy can be construed as a set of effective and acceptable courses of action implemented to reach explicit goals (Bridgman & Davis 2004). Implicit within this view is the assumption that policy makers are rational, though this assumption has been vigorously debated at times (Kingdon 1984, Stone 2002).

An early proponent of simplifying policy making by breaking it down to interrelated stages for the purpose of analysis was Lasswell (1956). This systemic analysis for understanding and explaining political systems served to convert inputs such as political demands and political support to outputs in the form of decisions and actions (Easton 1979). This idea was later extended to policies by Palmer (1997). Various attempts at classification of the different stages of the policy cycle have been carried out over the years.

**Figure #1 - The Policy Cycle**

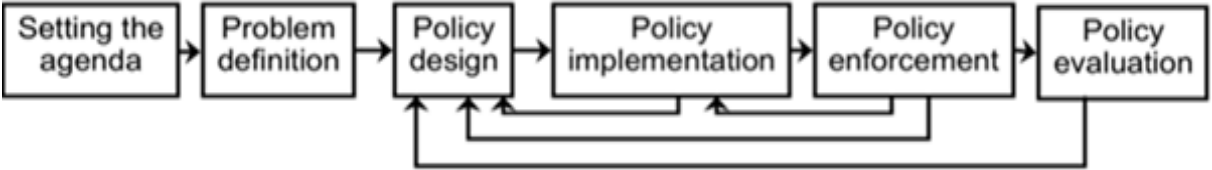

In this work, based upon the efforts of Stone (1988) and Howlett et al. (1995), the policy cycle is seen as a sequence of steps in which agenda setting; problem definition; policy design; policy implementation; policy enforcement; and policy evaluations are carried out in an iterative manner (see Figure #1).

**3.1 Policy Crowdsourcing Framework**
In this section, we combine the preceding analyses introduced thus far in the work to create an overarching Policy Crowdsourcing framework, including the different types of Crowdsourcing techniques merged with the various stages of the policy cycle. We then populate the resulting table (see Table#2) with the extant literature that we are aware of in the Policy Crowdsourcing domain, therein organizing, categorizing, and situating the extant research in detail.

**3.2 Framework Analysis**
As we inspect Tabke #2, it is immediately evident that the academic research on policy Crowdsourcing is relatively sparse, and a large portion of the potential space that might be covered is devoid of extant research. Of the sparse research that does exist, the vast majority is focused upon OC Crowdsourcing applications for policy purposes, while VLMs and TBCs have been relatively ignored. Perhaps not surprisingly, all of the research in respect to policy Crowdsourcing has emerged within the last five years, with most literature being much more recent than that. This seems to signal a growing application of Crowdsourcing for policy, given that most of the research is premised upon investigations of policy practitioner implementations.

As it stands at this point in time, OC holds the predominant share of research on policy Crowdsourcing, and this fact bears further fine-grained investigation. Is OC simply the Crowdsourcing application most suited for policy concerns? Or are extenuating factors in play?

**Table #2 – Policy Crowdsourcing Literature Situated in the different segments of the Policy Cycle**

|  | VLMs (E.G. M-Turk, Crowdflower, Gigwalk, Microworkers) | Tournaments (E.G. Innocentive, Kaggle, Eyeka, Challenge.gov) | Open Collaboration (E.G. Twitter, Reddit, Wikipedia, Ushahidi) |
|---|---|---|---|
| Agenda Setting |  | Brabham (2012b) Brabham (2013b) | Bott et al (2014) Aitamurto (2012) Charalabidis et al (2012) Bua (2012) Jungherr & Jürgens (2010) Lindner & Riehm (2011) |
| Problem Definition |  | Brabham (2013b) Basto et al. (2010) | Aitamurto (2012) Basto et al. (2010) Ferro et al (2013) |
| Policy Design |  | Brabham (2013b) Basto et al. (2010) | Nash (2009) Basto et al (2010) Bicquelet & Weale (2011) Seltzer & Mahmoudi (2012) Chun & Cho (2012) Stottlemyre & Stottlemyre (2012) Haklay et al (2014) |
| Policy Implementation |  | Brabham (2012) | Leeman et al (2014) Panagiotopoulos et al (2014) Brabham (2013) |
| Policy Enforcement | Kim et al (2013) |  | Bott et al (2014) Noveck (2009) Ghafele (2011) |
| Policy Evaluation |  | Basto et al. (2010) | Benkler et al (2013) Noveck (2009) Ghafele (2011) Gasser et al (2013) Ferro et al. (2013) Franklin et al (2013) Balagapo et al (2014) Schintler & Kulkarni (2014) |

In contrast, VLMs are barely represented in the research, and the reasons for this are not clear. It seems that the possible use of VLMs for policy has not caught-on among policymakers, leading to lower levels of academic research on the topic. Or it may be that VLMs are being used, but that this use is not well publicized, as opposed to TBCs and OC, which benefit from being publicized and are, by nature, more observable to outsiders. Further, it may be that ethics or national data sovereignty issues (Irion 2012), are more of a concern and therefore more of a hurdle to implement VLM's for policy. In any case, at the moment, VLMs are a largely untapped resource in this area. Similarly, TBCs are scarcely represented in the literature, which is somewhat surprising given the relatively high profile success of Challenge.gov (Brabham (2013b), and of Open Innovation platforms in non-policy domains in general).

In terms of the research methods employed in the literature that we reviewed, case studies (Yin 2014), and other phenomenon-based methods (Flick, 2002; Miles and Huberman, 1994) are the most common approaches to investigation and data collection that we have found. These methods have the benefit of being grounded in real-world conditions. However, they typically require considerable amounts of time to undertake, can be resource-intensive, and are generally limited in terms of generalizability. On the other hand, experiments (Trochim 2005) that utilize the various Crowds and applications available, have not been used to date. With the relatively low costs involved, and the low barriers to entry with Crowdsourcing, such experiments should be considered as a viable means to address the research gaps that we have identified in the application of Crowdsourcing to the policy cycle.

## 5. Conclusion

Our review of the extant literature on Crowdsourcing for policy has identified twenty-six pieces of literature that have recently emerged on the subject, which indicates that Crowdsourcing is already being used in different stages of the policy cycle, and that numerous researchers have deemed these efforts important and/or interesting enough to thoroughly study these instantiations. it may be that the existence of this extant literature may serve as a bellwether of sorts, indicating a new and rapidly emerging field of salient interdisciplinary social science inquiry.

Our work is an effort to coalesce this new field in a number of important directions which could serve to facilitate rapid growth in research and practice. In addition to collecting the extant research in one spot, we are situating and organizing this existing research by the general forms of Crowdsourcing. In doing so, we learn that research using all three forms of Crowdsourcing for policy already exists, though it is very far from evenly distributed across all the stages of the policy cycle. Along the same lines, we see that OC Crowdsourcing is by far the form of Crowdsourcing for policy most represented at this point in time, and the reasons for this are yet unclear, necessitating further investigation.

Similarly, our situation of the extant literature in the broader Crowdsourcing and policy cycle frameworks, readily indicates numerous gaps where no research exists at all. The reasons for this are also unclear, though it would seem to indicate that these voids represent useful research opportunities for future work. On the other hand, it may be that these apparent voids represent gaps in reality, indicating that the TBC's and VLM's may not yet have been used for Crowdsourcing those stages of the policy cycle. If this is at least partially true, then these gaps also represent opportunities for policy-makers and researchers alike to pioneer such efforts. Given the demonstrated value of all the Crowdsourcing techniques in other domains, it may be that these untapped potentials, severally or in combination, may represent powerful avenues to Crowdsource different stages of the policy cycle.

## References

AFUAH, A. & TUCCI, C. L. (2012) Crowdsourcing as a solution to distant search. Academy of Management Review. 37(3). p.355-375.11